%% file: main.tex
\documentclass{article}
\usepackage{arxiv_submission,times}

\input{math_commands.tex}

\usepackage{tikz}
\usetikzlibrary{positioning,arrows.meta}
\usepackage{hyperref}
\usepackage{url}
\usepackage{booktabs}
\usepackage{multirow}
\usepackage{graphicx}
\usepackage{threeparttable}
\usepackage{pgfplots}
\pgfplotsset{compat=1.18}
\title{Evaluator-in-the-Loop Monte Carlo Tree Search via LLM Agents for Motif Scaffolding in Protein Design}

\author{%
\begin{minipage}[t]{0.96\textwidth}
\raggedright
\textbf{Haotian Hu$^{1}$} \quad
\textbf{Oguzhan Gungordo$^{1}$} \quad
\textbf{Siheng Xiong$^{1}$} \quad
\textbf{Faramarz Fekri$^{1}$} \\
\vspace{0.4em}
\textnormal{$^{1}$Georgia Institute of Technology}
\end{minipage}
}

\begin{document}

\maketitle

\begin{abstract}
Motif-scaffolding systems commonly follow a generate-then-filter paradigm, in
which candidate proteins are generated independently and structural evaluation
is used primarily for terminal screening or ranking. This paradigm underuses
evaluation: failed predictions contain state-specific evidence about whether a
design requires repair of motif geometry, global foldability, or other
structural constraints.
We introduce \textbf{ELMS} (Evidence-based LLM-guided Monte Carlo Search), an
evaluator-in-the-loop search framework for motif scaffolding that turns such
evaluator feedback into targeted design actions. Effective reuse of structural feedback is nontrivial because
different scaffold states exhibit different failure modes, and repeatedly
refining a single trajectory can prematurely commit computation to an
unproductive region of sequence space. ELMS therefore retains evaluated
scaffolds as persistent search states: a Critic Agent diagnoses state-local
structural failures, a Policy Agent selects targeted operators with execution parameters,
motif-locked operators realize legal sequence modifications, and
MCTS determines which historical states should receive further design effort.
Under the standard GeomMotif protocol (100 candidates per task), ELMS achieves Successful rates
of 86.41\% on single-motif tasks and 84.57\% on paired-motif tasks, exceeding
the strongest prior baseline by 19.3 and 21.9 percentage points, respectively.
On MotifBench, under a matched 100-candidate search budget, it solves 26.7 of 30
tasks on average (88.89\% Task Success), compared with 16.0 tasks (53.33\%) for
the strongest baseline. These results establish ELMS as an effective
approach for converting structural evaluation from a terminal filter into
actionable guidance for iterative motif scaffolding.
\end{abstract}

\section{Introduction}

Motif scaffolding connects a local hypothesis of molecular function to a
physically realizable protein. A small set of residues may specify catalysis,
binding, or coordination through a precise three-dimensional arrangement, as
demonstrated in computational enzyme and epitope scaffolding
\citep{correia2014epitope}. The surrounding protein, however, must stabilize
this local geometry while forming a globally foldable and structurally coherent
scaffold, motivating structure-conditioned scaffold design approaches such as
constrained hallucination and inpainting \citep{wang2022scaffolding}. More
recently, diffusion-based protein design has enabled direct generation of
diverse scaffolds around prescribed structural motifs
\citep{watson2023rfdiffusion,butcher2025rfdiffusion3}, with subsequent methods extending this paradigm
toward more flexible and atomically specified motif constraints
\citep{ahern2026rfdiffusion2}. Despite these advances, motif scaffolding still
couples several objectives that need not improve together: preserving the
prescribed motif geometry, maintaining global foldability, and producing
structures sufficiently distinct from known or reference proteins. Reliably
balancing these local and global requirements therefore remains a central
challenge in functional protein design.

Most modern methods address motif scaffolding through a \emph{generate-then-filter}
paradigm. A motif-conditioned generator first produces a population of candidate
scaffolds, after which structure prediction and benchmark evaluation are used to
accept, reject, or rank the resulting designs. Diffusion-based approaches such
as RFdiffusion have made motif-conditioned backbone generation substantially more
effective \citep{watson2023rfdiffusion}, while flow-matching methods have
provided an alternative framework for sampling conditional protein structures
\citep{yim2024motifflow}. Subsequent methods such as Genie~2~\cite{lin2024genie2} and FADiff~\cite{liu2024fadiff} have
expanded the range and flexibility of geometric constraints that can be imposed
during generation. More recent approaches
extend motif-conditioned design toward finer-grained and atomistic specifications
\citep{ahern2026rfdiffusion2,geffner2026laproteina}. Despite these advances in
generation, structural evaluation is still typically consumed only after
candidates have been produced. Low structural confidence and motif displacement
indicate different failure modes and therefore suggest different corrective
actions, yet under a generate-then-filter interface they may both be treated
simply as unsuccessful samples. Consequently, evaluation is primarily used to
measure and filter failures rather than to determine how the next design should
be changed.

GeomMotif~\cite{strashnov2026geommotif} and MotifBench~\cite{zheng2025motifbench}
evaluate motif-scaffolding methods under benchmark-specified candidate
populations, measuring whether generated designs satisfy success criteria and,
in some analyses, novelty or diversity. Our question is complementary: when
structural evaluation, high-accuracy simulation, or expert review is costly, can
evaluator feedback be used during inference to discover at least one feasible
scaffold within a fixed candidate/evaluator budget more reliably than
generate-then-filter or simpler feedback-guided search?

To address this question, we introduce \textbf{E}vidence-based \textbf{L}LM-guided \textbf{M}onte Carlo \textbf{S}earch. ELMS
treats evaluated scaffolds as persistent search states: a Critic Agent summarizes
state-local structural failures, a Policy Agent selects a legal operator with execution parameters,
motif-locked operators realize the edit, and MCTS decides which
states receive further design effort. This turns structural evaluation from a
terminal filter into an inference-time signal for allocating a limited design
budget.

\begin{figure*}[t]
    \centering
    \includegraphics[width=0.98\textwidth]{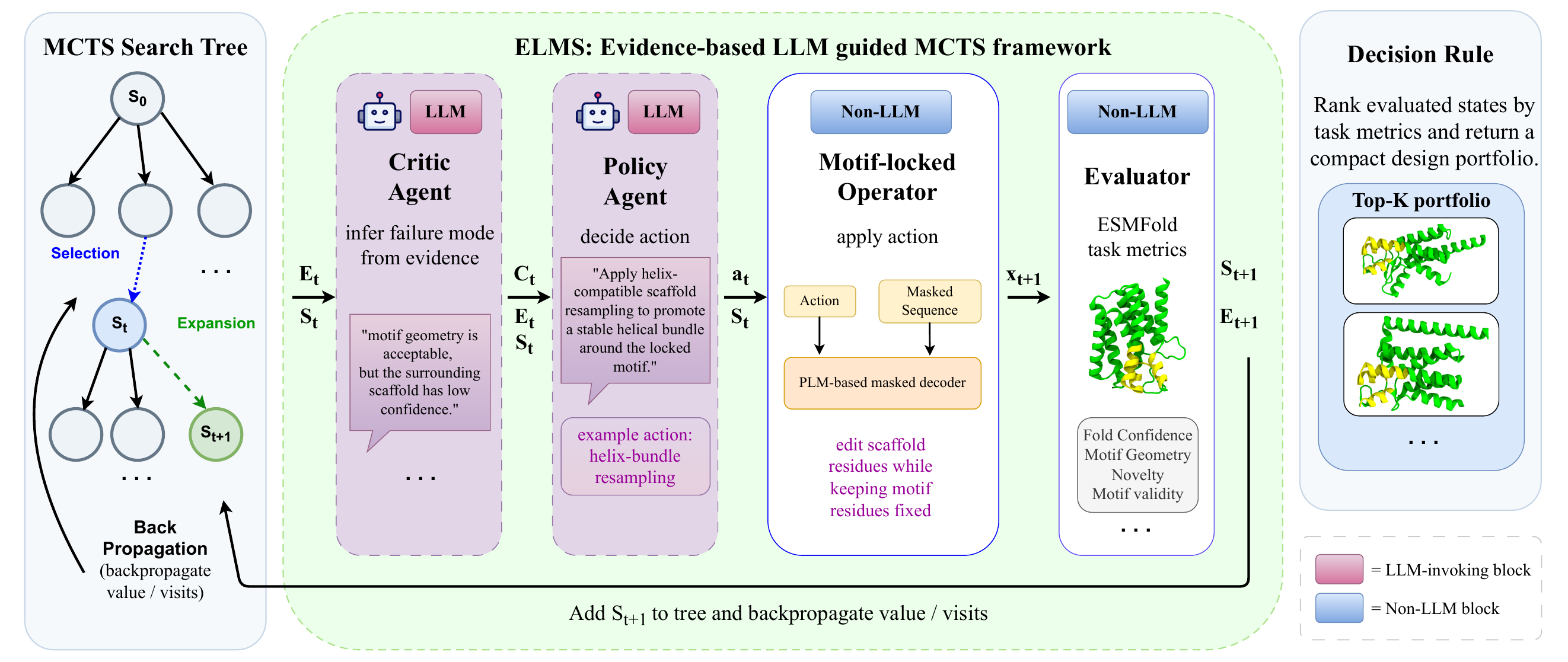}
    \caption{Overview of ELMS. The search tree selects an evaluated state for expansion; the Critic Agent analyzes the state-local evidence to identify which structural or task-specific constraints are limiting the current candidate, the Policy Agent translates this critic into a concrete edit plan over mutable sequence regions, and a motif-locked operator applies the edit while preserving the required motif positions to generate a new candidate sequence. Structural evaluation returns the
    resulting child to the search tree, where value and visit information is
    backpropagated to guide subsequent expansions. After search, evaluated states
    are ranked deterministically for benchmark reporting.}
    \label{fig:main_framework}
\end{figure*}

Our main contributions are:
\begin{itemize}
    \item We formulate motif scaffolding as ELMS, which conducts Monte Carlo Tree Search with an evaluator in the loop.
We therefore explicitly separate internal candidate-generation/evaluation budget from
    the final benchmark reporting interface.
    \item We introduce a multi-agent MCTS controller in which both LLM Critic and Policy Agents operate over state-local structural evidence, while motif-locked operators enforce
    constraints.
    \item We evaluate the framework on two complementary motif-scaffolding benchmarks. Under the standard GeomMotif Success protocol, it achieves Successful rates of 86.41\% on single-motif tasks and 84.57\% on paired-motif tasks. In a controlled MotifBench setting with 100 evaluated candidates per case, our framework solves 26.7 of 30 tasks on average, corresponding to 88.89\% Task Success. We further use matched-budget ablations and budget-scaling experiments to isolate the contribution of evaluator-guided tree search and analyze its sensitivity to inference compute.

\end{itemize}

\section{Related Work}

\textbf{Motif-conditioned protein generation.}
Computational motif scaffolding has progressed from physics-based scaffold
search to learned generative models capable of constructing proteins directly
around prescribed structural motifs. Classical Rosetta-based approaches placed
functional residues into existing or newly constructed scaffolds using
energy-based optimization and conformational sampling
\citep{kuhlman2003design,rothlisberger2008kemp,jiang2008denovo,correia2014epitope}.
Deep-learning approaches subsequently removed the need for predefined scaffold
libraries: constrained hallucination and RoseTTAFold inpainting generated
structures around functional sites \citep{wang2022scaffolding}, while
RFdiffusion established diffusion-based conditional backbone generation as a
strong motif-scaffolding paradigm \citep{watson2023rfdiffusion}. Later methods
expanded the flexibility of motif conditioning through SE(3) flow matching,
sequence diffusion, and multi-motif generation, including FrameFlow\cite{yim2024motifflow}, DPLM\cite{wang2024dplm},
Genie~2\cite{lin2024genie2}, and FADiff \citep{liu2024fadiff}.
More recently, RFdiffusion2 and La-Proteina have extended motif conditioning
toward unindexed and atom-level specifications
\citep{ahern2026rfdiffusion2,geffner2026laproteina}. Standardized benchmarks
such as MotifBench\cite{zheng2025motifbench} and GeomMotif\cite{strashnov2026geommotif} expose complementary limitations of these
models: MotifBench evaluates function-derived catalytic and binding motifs,
whereas GeomMotif isolates arbitrary geometric preservation across diverse
structural contexts.

\textbf{Evaluation-guided and agentic protein design.}
Modern protein-design pipelines commonly combine inverse folding and structure
prediction, using ProteinMPNN, AlphaFold, or ESMFold to test whether generated
sequences recover the intended structures
\citep{dauparas2022proteinmpnn,jumper2021alphafold,lin2023esmfold,hayes2025esm3}. Most
motif-scaffolding pipelines use these evaluations primarily for filtering,
ranking, or benchmark scoring. Related work has begun to use external feedback
more actively: guidance-based methods such as ProteinGuide steer pretrained
protein generators toward desired properties, while agentic systems such as
ProtAgents coordinate language models, scientific knowledge, and computational
tools during protein-design workflows \citep{xiong2025proteinguide,ghafarollahi2024protagents}.
These approaches demonstrate the value of incorporating additional evidence into
design, but generally treat feedback as a global conditioning signal or
tool-level reasoning resource rather than as candidate-specific evidence for
deciding how an evaluated scaffold should be edited next.

\textbf{Search and planning with evaluator feedback.}
Tree search methods provide a general mechanism for allocating inference-time
computation toward promising branches while preserving exploration
\citep{kocsis2006bandit,browne2012survey,xiong2026adaptive,gungordu2026pathwise,xiong2026enhancing}. In motif scaffolding, fold
confidence, motif geometry, and hard validity expose different online failure
modes, while novelty and redundancy are most naturally evaluated post hoc on the
accepted candidate set. Our work uses ELMS as a decision interface around fixed
structural evaluators. Unlike generate-then-filter pipelines, each evaluated
scaffold becomes a state whose evidence can guide the next expansion; unlike
unconstrained agentic design, LLM outputs are limited to critic summaries and
operator selection with execution parameters, while non-LLM modules enforce
motif locks, realize child sequences, evaluate structures, and back propagate values through the
tree.

\section{Preliminaries}

\subsection{Evaluation of Success and Novelty}
\label{sec:prelim_success_novelty}

We follow the evaluation protocol of GeomMotif~\citep{strashnov2026geommotif} and MotifBench~\cite{zheng2025motifbench} to assess whether a generated scaffold both preserves the target motif geometry and departs from the reference fold.
\paragraph{Success.}
We use benchmark-specific success definitions. In GeomMotif, we use the benchmark's modality-specific success
criteria. For our sequence-generated candidates, success requires
motif-residue $C_\alpha$ RMSD $\leq 1.0\,\text{\AA}$ and mean pLDDT
$\geq 70$. Structure-generator baselines retain the benchmark's
native structure-level evaluation protocol, which uses the motif $C_\alpha$
RMSD gate together with a structure-level self-consistency gate rather than
pLDDT.
In MotifBench, success is defined at the submitted
backbone level: each submitted backbone is sequence-designed eight times with
ProteinMPNN and refolded with ESMFold, and the backbone is successful if at least
one refolded design satisfies both motif RMSD $<1.0~\text{\AA}$
over motif-residue N, $C_\alpha$, and C atoms and backbone self-consistency
RMSD $<2.0~\text{\AA}$ over backbone $C_\alpha$ atoms. Following
MotifBench, no pLDDT or confidence filter is applied to the final
backbone-success decision.
\paragraph{Novelty.}
Success alone does not distinguish genuinely new scaffolds from candidates that simply reconstruct the original reference fold. Therefore, novelty is evaluated among successful candidates by comparing each generated scaffold to the corresponding ground-truth structure using TM-score. We denote the ground-truth reference scaffold for the current task as $S_{\mathrm{ref}}$. A successful candidate is treated as novel when its TM-score to the reference structure is below $0.8$:
\[
\mathrm{Novel}(S_t)
=
\mathbb{I}\left[
\mathrm{Success}(S_t)=1
\;\land\;
\mathrm{TMscore}(S_t, S_{\mathrm{ref}}) \leq 0.8
\right].
\]
\subsection{Action Library and Motif-Locked Operators}
\label{sec:prelim_action_library}

We use a fixed LLM-synthesized library of biologically motivated
motif-locked operators, distilled from protein-design literature,
frozen before evaluation, and reused unchanged across GeomMotif
and MotifBench. During search, the Policy Agent selects an operator
identifier and execution parameters. Actions are realized through
fixed-length masked decoding with a protein language model.
Decoding may be stochastic, but modifies only permitted scaffold
positions while preserving prescribed motif residue identities
and their assigned sequence positions. Folded motif geometry is
assessed separately by the structure evaluator. The operator
inventory, routing, and decoding details are provided in
Appendix.
\input{methodology_mcts}

\section{Experiments}
\label{sec:experiments}

\subsection{Experimental Questions and Reporting Policy}
\label{subsec:experiment_reporting_policy}

The experiments address two related questions. First, under the standard
GeomMotif protocol, can our framework improve benchmark Successful
rates while using the original candidate protocol and success definition? Second,
under a fixed candidate/evaluator budget, can ELMS discover successful scaffolds
more reliably than non-adaptive generation or simpler iterative feedback search?
Benchmark splits, candidate interfaces, and baseline-number sources are
summarized in Appendix.

Success is the primary main-text endpoint because it is the fixed
benchmark-level feasibility question. During search, ELMS uses only
candidate-local evaluator feedback and online proxy signals; final benchmark
success is recomputed under each benchmark protocol after candidate generation.
Novelty is reported post hoc: it is evaluated after the final
candidate set is formed and is therefore treated as a characterization metric
rather than primary online objectives.

\subsection{GeomMotif Success and Novelty Comparison}
\label{subsec:geommotif_success_results}

Table~\ref{tab:geommotif_results} compares our framework with published
GeomMotif baselines using the original GeomMotif Successful definition and
candidate protocol (100 candidates per task), while reporting Novelty as a secondary post hoc statistic.
Protocol details are provided in Appendix.
ELMS reaches 86.41\% Successful on single-motif tasks and 84.57\% on
paired-motif tasks, while also reaching 83.3\% Novel on single-motif tasks and
65.7\% Novel on paired-motif tasks. These gains substantially exceed the
strongest baseline methods on both single- and paired-motif tasks, indicating
that evaluator-guided search improves feasibility across GeomMotif constraint
settings. We also include a \emph{Baseline LLM} that directly revises the current sequence
from candidate-local evaluator feedback, without the action library, branch
preservation, UCB selection, or value back propagation.

\begin{table*}[t]
\centering
\caption{GeomMotif success and novelty comparison. All rows use the standard
GeomMotif protocol (100 candidates per task) and benchmark Successful definition. Values are
reported as mean $\pm$ standard deviation. Post hoc Unique and additional diversity statistics are reported in
Appendix.}
\label{tab:geommotif_results}
\begin{threeparttable}
\small
\setlength{\tabcolsep}{7pt}
\renewcommand{\arraystretch}{1.05}
\centering
\begin{tabular}{@{}lcccc@{}}
\toprule
\multirow{2}{*}{Model / system}
& \multicolumn{2}{c}{Successful (\%)}
& \multicolumn{2}{c}{Novel (\%)} \\
\cmidrule(lr){2-3}
\cmidrule(lr){4-5}
& Single & Paired & Single & Paired \\
\midrule
Genie2
& $60.1 \pm 1.0$ & $32.9 \pm 0.4$ & $60.1 \pm 1.0$ & $26.6 \pm 0.5$ \\
La-Proteina
& $67.1 \pm 0.7$ & $62.7 \pm 0.5$ & $67.1 \pm 0.6$ & $35.2 \pm 0.9$ \\
RFdiffusion
& $65.1 \pm 0.4$ & $43.7 \pm 1.0$ & $65.1 \pm 0.4$ & $25.0 \pm 0.5$ \\
Protpardelle-1C
& $56.2 \pm 0.7$ & $44.6 \pm 0.5$ & $56.2 \pm 0.7$ & $25.2 \pm 0.4$ \\
FrameFlow
& $30.6 \pm 0.4$ & $25.1 \pm 0.4$ & $30.6 \pm 0.4$ & $19.7 \pm 0.3$ \\
RFdiffusion2
& $24.9 \pm 0.6$ & $13.5 \pm 0.5$ & $24.9 \pm 0.6$ & $11.4 \pm 0.3$ \\
ESM3 (seq)
& $17.4 \pm 0.3$ & $6.5 \pm 0.5$ & $11.3 \pm 0.5$ & $0.1 \pm 0.0$ \\
DPLM-3B
& $19.3 \pm 0.7$ & $11.0 \pm 0.5$ & $10.2 \pm 0.5$ & $0.9 \pm 0.2$ \\
DPLM-650M
& $15.9 \pm 0.4$ & $6.5 \pm 0.5$ & $8.7 \pm 0.2$ & $0.4 \pm 0.1$ \\
ESM3 (seq \& struct)
& $2.0 \pm 0.1$ & $0.7 \pm 0.2$ & $2.0 \pm 0.1$ & $0.0 \pm 0.0$ \\
\addlinespace
Baseline LLM (gpt-5-nano)
& $13.1 \pm 0.1$ & $7.2 \pm 0.2$ & $3.7 \pm 0.0$ & $0.3 \pm 0.0$ \\
\textbf{ELMS} (gpt-5-nano)
& $\mathbf{86.4}\pm 0.8$ & $\mathbf{84.6}\pm 0.6$ & $\mathbf{83.3}\pm 0.5$ & $\mathbf{65.7}\pm 0.6$ \\
\bottomrule
\end{tabular}%
\end{threeparttable}
\end{table*}

\subsection{MotifBench De Novo Design Comparison}
\label{subsec:motifbench_results}

MotifBench contains 30 functional motif-scaffolding tasks. We compare methods
under a matched 100-candidate search budget per task. Candidates are ranked
deterministically for benchmark reporting; because this ranking uses the
benchmark success indicator, Task Success is interpreted as whether the
100-candidate pool contains at least one successful candidate. Additional
protocol details are provided in
Appendix.

Table~\ref{tab:motifbench_results} reports the fixed-search-budget MotifBench comparison. Across three runs, \textsc{ELMS} solves 26.7 of 30 tasks on average and achieves an 88.89\% Success Rate, exceeding the strongest baseline by 10.7 solved tasks and 35.56 percentage points (16.0 tasks, 53.33\%). This large gain shows that evaluator-in-the-loop LLM-guided search is effective at using a fixed search budget to discover successful scaffolds.

\begin{table*}[t]
\centering
\caption{MotifBench fixed-search-budget comparison. Results are averaged over
three independent runs; each run evaluates 100 candidates per task. Solutions
denotes the mean number of tasks for which at least one evaluated candidate
satisfies the MotifBench success criterion.}
\label{tab:motifbench_results}
\begin{threeparttable}
\setlength{\tabcolsep}{12pt}
\renewcommand{\arraystretch}{1.08}
\centering
\begin{tabular}{@{}lcc@{}}
\toprule
Method & Solutions &  Success Rate(\%) \\
\midrule
RFdiffusion3
& 15.7 & $52.22 \pm 4.84$ \\
La-Proteina
& 16.0 & $53.33 \pm 1.92$ \\
Protpardelle-1c (cc91-large)
& 14.7 & $48.89 \pm 1.11$ \\
Genie3
& 9.0 & $30.00 \pm 1.92$ \\
Baseline LLM(GPT-5-nano)
& 8.3 & 27.78 $\pm$ 2.22 \\
\textbf{ELMS}(GPT-5-nano)
& \textbf{26.7} & $\mathbf{88.89 \pm 1.11}$ \\
\bottomrule
\end{tabular}

\end{threeparttable}
\end{table*}

\subsection{Matched-Budget Search Ablation}
\label{subsec:search_ablation}

This ablation asks whether the observed GeomMotif Success improvement comes from
ELMS rather than from changing the underlying language
model. All variants are evaluated on GeomMotif and use the same LLM model
(gpt-5-nano), the same blinded motif inputs, the same legal action space, the
same structural evaluator, the same Success gate, and the same total
candidate/evaluator budget $B_{\mathrm{search}}=100$.

We instantiate four matched-budget search variants. \emph{Random Operator Search} samples randomly from the shared legal action library without using evaluator feedback for routing. \emph{Generate-then-filter} independently generates candidates without feeding evaluator results back into action selection. \emph{Greedy Operator Search} uses candidate-local evaluator feedback to select operators, but follows only the current trajectory, with no tree selection or value backpropagation. \emph{ELMS} uses the same feedback and action library while preserving multiple branches, applying UCB selection, and back propagating evaluator-derived values. Unlike Baseline LLM, Greedy Operator Search is therefore a matched-library search framework, differing from ELMS only in its tree-level search machinery rather than its action-selection interface. In Table~\ref{tab:search_ablation}, ``Evaluator feedback'' indicates
whether later action choices can condition on structural evaluator outputs, and
``Tree selection / value backpropagation'' describes how evaluated feedback is retained and reused.

\begin{table*}[t]
\centering
\caption{Matched-budget GeomMotif search ablation with identical
LLM, inputs, evaluator, success gate, and action space, using
$B_{\mathrm{search}}=100$ evaluated candidates per task.
All rows report Task Success: the percentage of tasks with
at least one successful candidate.}
\label{tab:search_ablation}
\setlength{\tabcolsep}{7pt}
\renewcommand{\arraystretch}{1.08}
\resizebox{\linewidth}{!}{%
\begin{tabular}{@{}lccccc@{}}
\toprule
Variant & Evaluator feedback & Action routing
& Tree selection / value backpropagation & Single Success (\%) & Paired Success (\%) \\
\midrule
Random Operator Search & No & Random legal routing & No
& 0.0 & 0.0 \\
Generate-then-filter & No & Independent direct generation & No
& 6.1 & 10.4 \\
Greedy Operator Search & Yes & Evidence-guided greedy routing & No; single trajectory
& 48.0 & 53.7 \\
ELMS & Yes & Evidence-guided action selection
& UCB selection + value backpropagation
& \textbf{88.6} & \textbf{86.4} \\
\bottomrule
\end{tabular}%
}
\end{table*}

\subsection{Search-Budget Scaling}
\label{subsec:budget_scaling}

We next examine how the two feedback-guided search methods use additional inference compute. We fix $K_{\mathrm{sub}}=1$ and vary the internal search budget over $B_{\mathrm{search}}\in\{{5,10,20,50,100}\}$. Task Success@1 is the fraction of 57 GeomMotif tasks for which the returned design satisfies the Success gate. Greedy Operator Search and ELMS share the same blinded inputs, evaluator, and legal action space, differing only in whether evaluated alternatives are retained and exploited through tree statistics and value backpropagation.

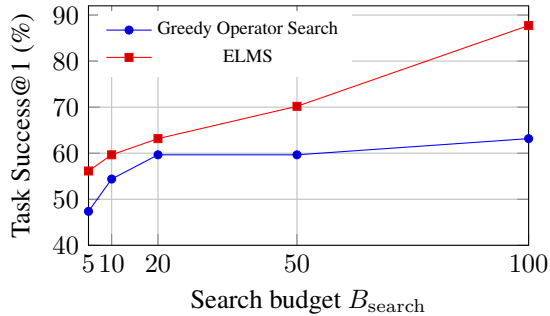
\begin{figure}[t]
\centering
\begin{tikzpicture}
\begin{axis}[
    width=0.53\linewidth,
    height=0.34\linewidth,
    xlabel={Search budget $B_{\mathrm{search}}$},
    ylabel={Task Success@1 (\%)},
    xmin=5, xmax=100,
    ymin=40, ymax=92,
    xtick={5,10,20,50,100},
    ytick={40,50,60,70,80,90},
    grid=major,
legend style={
    draw=none,
    at={(0.02,0.98)},
    anchor=north west,
    font=\scriptsize
},    mark size=1.5pt
]
\addplot+[mark=*] coordinates {
    (5,47.37) (10,54.39) (20,59.65) (50,59.65) (100,63.16)
};
\addlegendentry{Greedy Operator Search}
\addplot+[mark=square*] coordinates {
    (5,56.14) (10,59.65) (20,63.16) (50,70.18) (100,87.72)
};
\addlegendentry{ELMS}
\end{axis}
\end{tikzpicture}
\caption{GeomMotif Task Success@1 as a function of internal search budget with
$K_{\mathrm{sub}}=1$. Greedy feedback improves at small budgets but largely
plateaus, whereas ELMS continues to benefit from additional search.}
\label{fig:budget_scaling}
\end{figure}

Figure~\ref{fig:budget_scaling} shows different scaling behavior for the two
controllers. Greedy Operator Search rises from 47.37\% at $B=5$ to 59.65\% at
$B=20$, but reaches only 63.16\% at $B=100$. ELMS rises from 56.14\% to
87.72\% over the same budget range. In particular, from $B=20$ to $B=100$,
greedy search improves by only 3.51 percentage points, while MCTS improves by
24.56 points. The widening gap indicates that the benefit of tree search is not
simply due to receiving evaluator feedback: preserving evaluated alternatives
allows additional inference compute to continue producing useful exploration
rather than repeatedly refining a single trajectory.

\section{Conclusion}
\label{sec:conclusion}

We introduced ELMS for motif scaffolding, which turns structural evaluation into state-specific
guidance for iterative protein design. ELMS retains evaluated
scaffolds as persistent search states, uses Critic and Policy Agents to
interpret structural evidence and select targeted modification strategies,
executes these decisions through motif-locked operators, and
uses MCTS to adaptively allocate further design effort across promising and
repairable states.

Across GeomMotif and MotifBench, the framework consistently outperforms the strongest
generative baselines under matched candidate/evaluator budgets. These results highlight ELMS as a promising approach for combining structural evaluation, agentic reasoning, and adaptive
search in constrained protein design. Future work may incorporate richer
structural and functional evidence, extend the action space beyond
sequence-level refinement, and evaluate whether evaluator-guided search can
generalize to broader protein-design objectives and experimentally validated
design settings.

\section*{AI Use Statement}

Generative AI tools used for manuscript preparation were used solely to improve the clarity, grammar, and readability of the manuscript. They were not used for research ideation, methodology or experimental design, implementation, or result interpretation. Representative examples of wording-only edits are provided in Appendix.

\bibliographystyle{plainnat}
\bibliography{references}

\end{document}

%% file: math_commands.tex
\usepackage{amsmath,amsfonts,bm}

\def\eqref#1{equation~\ref{#1}}

\def\1{\bm{1}}

\DeclareMathAlphabet{\mathsfit}{\encodingdefault}{\sfdefault}{m}{sl}
\SetMathAlphabet{\mathsfit}{bold}{\encodingdefault}{\sfdefault}{bx}{n}



%% file: methodology_mcts.tex
\section{Methodology}
\label{sec:methodology}
\subsection{Framework Overview}
ELMS improves motif scaffolding by using LLM agents to transform state-local structural feedback into targeted intervention decisions, enforcing motif preservation through motif-locked sequence operators, and adaptively allocating search effort toward promising or repairable design trajectories.

We formulate the search as a bounded MCTS problem
\citep{kocsis2006bandit,browne2012survey}. Each tree node stores an evaluated
motif-constrained scaffold state
\[
S_i=(x_i,\Pi_i,\boldsymbol{\ell}_i,p_i,a_i^{\mathrm{in}},E_i,N_i,W_i,Q_i).
\]
Here, $i$ indexes a node in the MCTS tree. $x_i \in \mathcal{A}^{L_i}$ is the
protein sequence over amino-acid alphabet $\mathcal{A}$, and $L_i$ is its total
sequence length. $\boldsymbol{\ell}_i$ records the scaffold lengths around the
motif segments: if a task contains $m$ ordered motif segments, then
\[
\boldsymbol{\ell}_i=(\ell_{i,0},\ell_{i,1},\ldots,\ell_{i,m})
\]
records the length before the first motif segment, the lengths between adjacent
motif segments, and the length after the last motif segment. Together with the
fixed motif segment lengths, these values determine $L_i$. $\Pi_i$ is the
motif-to-scaffold placement map, which maps each motif residue to a position in
$\{1,\ldots,L_i\}$ while satisfying benchmark placement constraints and
preserving motif order. The parent pointer $p_i$ denotes the parent node of
$S_i$, and $a_i^{\mathrm{in}}$ records the incoming motif-locked action
specification that generated $S_i$ from that parent. $E_i$ is the cached structural evidence obtained when $S_i$ is
created, including motif RMSD, pLDDT, hard-validity signals, and online feasibility-proxy
margins. $(N_i,W_i,Q_i)$ are the visit count, accumulated value, and mean
value used by MCTS.

As shown in Fig.~\ref{fig:main_framework}, each iteration follows a
select--critique--choose-action--apply--evaluate--backpropagate loop. Here,
superscript $(r)$ indexes the current expansion iteration and is separate from
the persistent node index $i$. The selected parent state is
$S_{\mathrm{par}}^{(r)}$, and the newly generated child state is
$S_{\mathrm{child}}^{(r)}$. The Critic Agent reads the cached evidence
$E_{\mathrm{par}}^{(r)}$ associated with $S_{\mathrm{par}}^{(r)}$ and produces
a critique $C^{(r)}$. Conditioned on
$(S_{\mathrm{par}}^{(r)},E_{\mathrm{par}}^{(r)},C^{(r)})$, the Policy Agent
returns a legal action specification $a^{(r)}$ from the operator library in
Appendix. This specification includes an operator
identifier and execution parameters such as search scale, mutable positions,
repair fraction, and sampling temperature. The corresponding motif-locked
operator applies the action to legal scaffold positions while preserving the
prescribed motif residues, producing $S_{\mathrm{child}}^{(r)}$. The child state is then folded
and evaluated to obtain $E_{\mathrm{child}}^{(r)}$, and the resulting search utility is
backpropagated to update the MCTS statistics.
\[
\begin{tikzpicture}[
    baseline=(tree.base),
    node distance=1.45cm,
    every node/.style={inner sep=1pt},
    >={Stealth[length=2mm]}
]
\node (tree) {$\mathrm{MCTS\ Tree}$};
\node[right=of tree] (state) {$S_{\mathrm{par}}^{(r)}, E_{\mathrm{par}}^{(r)}$};
\node[right=of state] (critic) {$C^{(r)}$};
\node[right=of critic] (action) {$a^{(r)}$};
\node[right=of action] (next) {$S_{\mathrm{child}}^{(r)}$};
\node[right=of next] (eval) {$E_{\mathrm{child}}^{(r)}$};

\draw[->] (tree) -- node[above] {\scriptsize select} (state);
\draw[->] (state) -- node[above] {\scriptsize Critique} (critic);
\draw[->] (critic) -- node[above] {\scriptsize Select action} (action);
\draw[->] (action) -- node[above] {\scriptsize Apply} (next);
\draw[->] (next) -- node[above] {\scriptsize Evaluate} (eval);

\draw[->] (eval.south) 
    -- ++(0,-0.6cm)
    -- node[above] {\scriptsize backpropagation} ([yshift=-0.6cm]tree.south)
    -- (tree.south);
\end{tikzpicture}
\]

\subsection{Search and Submission Budgets}
\label{sec:decision_problem}

Let $B_{\mathrm{search}}$ denote the internal search budget, defined as the number of candidate states evaluated by the structural prediction and evaluation pipeline. Let $K_{\mathrm{sub}}$ denote the number of final designs retained for benchmark reporting. These two quantities correspond to different practical stages: the search procedure may evaluate multiple intermediate states while exploring and repairing design trajectories, whereas a benchmark or downstream workflow may retain only a smaller reported set. We use this distinction to separate internal search effort from the final reporting interface when analyzing budget sensitivity. When the terminal ranking uses the benchmark success indicator, task solving at retained size $K_{\mathrm{sub}}$ is interpreted as whether the evaluated pool contains a successful candidate under $B_{\mathrm{search}}$, not as an independent blind Top-$K_{\mathrm{sub}}$ selection problem.

\subsection{LLM Agents and Non-LLM Modules}
\label{sec:agent_module_map}

Table~\ref{tab:agent_module_map} separates the two LLM components, the Critic
and Policy Agents, from non-LLM code and fixed pretrained protein-model
inference. Both agents use gpt-5-nano with constrained JSON outputs; all numeric
measurements, constraint checks, tree statistics, sequence edits, refolding
calls, and benchmark metrics are produced outside the LLM.
\begin{table*}[t]
\centering
\caption{Role map for ELMS. The table identifies which roles are dedicated
LLM agents and which ones are non-LLM modules or fixed protein-model calls.}
\label{tab:agent_module_map}
\small
\resizebox{\linewidth}{!}{%
\begin{tabular}{@{}llll@{}}
\toprule
Role & Implementation & LLM? & Output \\
\midrule
MCTS search loop & Deterministic UCB/back propagate functions & No
& Selected tree node, visit counts, back propagate values \\
Critic Agent & gpt-5-nano with ontology-constrained JSON & Yes
& Candidate-local critic labels and rationales \\
Policy Agent & gpt-5-nano with action-schema-constrained JSON & Yes
& Legal operator choices plus execution parameters \\
Motif-locked operator & Motif-lock template plus sequence-model backend & No
& Apply selected actions to legal scaffold regions while preserving motif residues \\
Structure evaluator & ESMFold plus benchmark-specific adapters & No
& Fold confidence, motif RMSD, hard-validity signals, online proxy status \\
Benchmark reporter & Deterministic ranking procedure & No
& Ranked evaluated candidate set \\
\bottomrule
\end{tabular}%
}
\end{table*}

This design prevents the LLM from inventing benchmark scores or bypassing motif
constraints. The agents produce only constrained critic labels and legal
action-schema outputs, including operator identifiers and execution parameters;
motif-locked operators then apply the selected edits only to legal
scaffold regions while preserving prescribed motif residues; motif-lock and
validity checks are enforced deterministically. All numeric
measurements and benchmark metrics come from the structure evaluator, not the
LLM. The no-leakage interfaces, prompts, and schema validation rules are provided
in Appendix.

\subsection{Task Representation and Blinded State Initialization}

A benchmark instance is represented as
\[
\mathcal{T}=(M,C,L,Ref),
\]
where $M$ specifies public motif residues and fixed identities, $C$ contains
placement or contig constraints, $L$ is the allowed length range and $Ref$ contains
benchmark-provided reference information used only by the evaluator. The framework sees only the
public motif mask, public motif identities, generated sequences, predicted
structures, and candidate-local structural feedback.
For both GeomMotif and MotifBench, each task is initialized from benchmark-provided motif constraints rather than from a benchmark-origin scaffold sequence. The task adapter reads the required motif specification, including motif residues or segments, target length or allowed length range, motif-placement information, and reference motif coordinates. The initial state $S_0$ is a virtual motif-only root that stores these sanitized constraints but is not itself structurally evaluated. Root expansion uses legal restart or budgeted sequence-only operators to sample complete scaffold sequences by masking all non-motif scaffold positions under the chosen length and placement constraints. These operators preserve the prescribed motif residues, modify only legal scaffold regions, and provide broad sequence initializations before any evaluated parent exists.

After the first complete candidates are generated, they become the first evaluated tree nodes. All subsequent states are produced by MCTS expansion, motif-locked sequence modification, and fresh structural evaluation under the corresponding benchmark adapter. 

\subsection{MCTS Selection, Expansion, Evaluation, and Backpropagation}
\label{sec:mcts_loop}

At iteration $r$, the search procedure descends the current tree using an
upper-confidence criterion. For an already evaluated child node $j$ of parent
node $i$ with $N_j>0$, the default selection score is
\[
\mathrm{UCB}(i,j)=Q_j+c\sqrt{\frac{\log(N_i+1)}{N_j+1}}.
\]
Here, $N_i$ is the number of backed-up child evaluations whose selected path
contains node $i$, and $N_j$ is the corresponding count for child node $j$.
Thus, if a batch expansion evaluates multiple children with a shared ancestor,
that ancestor receives one visit update per evaluated child. $Q_j$ is defined
only after $N_j>0$ and equals the accumulated backpropagated value divided by
$N_j$. Legal but unevaluated children are expanded before this score is computed;
the $+1$ terms stabilize exploration among visited nodes and do not define
$Q_j$ for an unvisited child. The exploration constant $c$ controls how strongly
the search favors less-visited children.
Among visited children, deterministic tie-breaking is performed by depth,
hard-validity, and evaluator margin before repeatedly exploiting the current
best-valued branch.

Expansion converts an evaluated leaf into one or more legal action specifications.
The Critic Agent receives a compact node record containing the sequence
length, public motif mask, parent action, current online metrics, evidence
records, and visit statistics. It emits ontology labels such as
\texttt{quality\_failed}, \texttt{motif\_geometry\_failed},
\texttt{near\_threshold}, or \texttt{state\_successful}. The Policy Agent then
maps the verified critic to legal operator choices and execution parameters from
the operator library.

Each accepted action is realized by the motif-locked operator backend rather than by the
LLM. The backend constructs an editable mask over non-motif residues, preserves
motif identities and motif-to-sequence mapping, validates the amino-acid
alphabet and length constraints, removes duplicates, and only then admits the
child for evaluation. Newly generated sequences are folded with ESMFold
\citep{lin2023esmfold}; benchmark adapters compute fold confidence, motif RMSD,
online feasibility-proxy margins, and hard-constraint flags. Final benchmark
success, including MotifBench backbone-level success, is recomputed separately
under the benchmark protocol.

After evaluation, the newly evaluated child node $j$ receives a scalar search
utility $R_j$, computed from its state-local evaluation record as detailed in
Appendix. This value is then backpropagated to every
ancestor node $i$ on the selected path:
\[
N_i \leftarrow N_i+1,
\qquad
W_i \leftarrow W_i+R_j,
\qquad
Q_i \leftarrow W_i/N_i.
\]
The backpropagated value affects future tree selection only. The individual
metrics remain available to the Critic Agent, Policy Agent, and terminal
selector so that high-confidence but geometrically incorrect candidates are not
collapsed into the same explanation as low-confidence but geometry-preserving
candidates.

\subsection{Termination and Benchmark-Level Reporting}

Iteration in Fig.~\ref{fig:main_framework} stops when the tree reaches the
evaluator budget or when no legal children remain.
Benchmark reporting uses only structurally evaluated states. For each
task $t$, let $\mathcal{V}_t$ denote the set of evaluated states in the MCTS
tree. We rank states in $\mathcal{V}_t$ by a fixed deterministic ordering:
hard-valid candidates are preferred first, followed by benchmark success, lower
motif RMSD and higher fold confidence. When a benchmark requires a retained
set, $\mathcal{P}_t^{K_{\mathrm{sub}}} \subseteq \mathcal{V}_t$ denotes the
top-$K_{\mathrm{sub}}$ states under this ordering; if fewer than
$K_{\mathrm{sub}}$ evaluated states exist, all evaluated states are retained.
An empty evaluated pool or retained set contributes zero to the task-level
indicator. The primary task-solving endpoint is
\[
\mathrm{TaskSolved}(B_{\mathrm{search}})=
\frac{1}{|\mathcal{T}|}\sum_{t\in\mathcal{T}}
\mathbf{1}\left[\exists i\in\mathcal{V}_t:\; y_i=1\right],
\]
where $y_i \in \{0,1\}$ is the benchmark-specific success indicator for
evaluated state $i$. Because success-aware terminal ranking places successful
candidates before failures, task solving at retained size $K_{\mathrm{sub}}$ is
equivalent to this pool-level quantity for any $K_{\mathrm{sub}}\geq1$. Task
Success denotes this task-level quantity, whereas $y_i$ remains the
candidate-level success indicator.

Benchmark-level comparisons follow the evaluation protocols of the corresponding benchmarks. For GeomMotif, we report the standard Successful rate using the original benchmark success gate. For MotifBench, we use Task Success as a fixed-search-budget endpoint under $B_{\mathrm{search}}=100$ evaluated candidates per task. Additional protocol details, auxiliary benchmark metrics, and clustering-based analyses are reported in Appendix.

\makeatletter
\let\mcts@orig@tabular\tabular
\let\mcts@orig@endtabular\endtabular
\renewenvironment{tabular}[1]{%
  \ifnum\value{table}=4\relax
    \scriptsize
    \setlength{\tabcolsep}{3pt}%
  \fi
  \mcts@orig@tabular{#1}%
}{%
  \mcts@orig@endtabular
}
\makeatother